\documentclass{article}
\usepackage{natbib,epsfig}

\usepackage{amsthm,amsmath,amssymb,colonequals,multirow,hyperref,graphicx,epstopdf,enumerate}
\newcommand{\vecxc}{\mathbf{X}}

\newcommand{\vecu}{\mathbf{u}}
\newcommand{\vecx}{\mathbf{x}}
\newcommand{\vecz}{\mathbf{z}}

\newcommand{\vecmu}{\mbox{\boldmath$\mu$}}

\newcommand{\matsig}{\mathbf{\Sigma}}

\newcommand{\varthet}{\boldsymbol{\vartheta}}
\newcommand{\vecthet}{\mbox{\boldmath$\vartheta$}}
\newenvironment{theorem}[1][Theorem]{\begin{trivlist}
\item[\hskip \labelsep {\bfseries #1}]}{\end{trivlist}}
\newenvironment{pta}[1][Part a]{\begin{trivlist}
\item[\hskip \labelsep {\bfseries #1}]}{\end{trivlist}}
\newenvironment{ptb}[1][Part b]{\begin{trivlist}
\item[\hskip \labelsep {\bfseries #1}]}{\end{trivlist}}

\begin{document}

\title{A LASSO-Penalized BIC for Mixture Model Selection}

\author{Sakyajit Bhattacharya  and~Paul~D.~McNicholas\thanks{Department of Mathematics \& Statistics, University of Guelph, Guelph, Ontario, N1G 2W1, Canada. E-mail: paul.mcnicholas@uoguelph.ca.}% <-this % stops a space
}
\date{Department of Mathematics \& Statistics, University of Guelph.}
\maketitle

\begin{abstract}
The efficacy of family-based approaches to mixture model-based clustering and classification depends on the selection of parsimonious models. 
%For example, mixtures of factor analyzers  and extensions thereof are developing into effective approaches for the analysis of high-dimensional data. 
Current wisdom suggests the Bayesian information criterion (BIC) for mixture model selection. However, the BIC has well-known limitations, including a tendency to overestimate the number of components as well as a proclivity for, often drastically, underestimating the number of components in higher dimensions. While the former problem might be soluble through merging components, the latter is impossible to mitigate in clustering and classification applications. In this paper, a LASSO-penalized BIC (LPBIC) is introduced to overcome this problem. This approach is illustrated based on applications of extensions of mixtures of factor analyzers, where the LPBIC is used to select both the number of components and the number of latent factors. The LPBIC is shown to match or outperform the BIC in several situations. 
\end{abstract}

%\begin{table}
%\caption{\label{tab:simst}Mean parameter estimates from the application of our mixture of generalized hyperbolic distribution to 100 simulated data sets from a two-component mixture of skew-$t$ distributions.}
%\centering
%\fbox{
%\begin{tabular}{*{6}{r}}
%& \multicolumn{2}{c}{$g=1$} & & \multicolumn{2}{c}{$g=2$}\\
% \cline{2-3}\cline{5-6}
% & True  & Estimated &  & True & Estimated\\
%\hline
%$\vecmu_g$ & $(3.00,3.00)$ & $(2.95,3.04)$ &&$(-3.00,-3.00)$ & $(-2.89,-3.11)$\\
%$\vecalpha_g$ &$(2.00,-2.00)$ & $(2.05,-2.06)$&& $(-1.00,1.00)$ &$(-1.30,1.36)$\\
%$\matsig_g$ & $(1.00,-0.75,1.00)$ &$(0.99,-0.74,0.98)$ &&$(1.00,-0.75,1.00)$ & $(1.01,-0.76,1.01)$\\
%$\omega_g$ & $0.00$& $0.00$&  &$0.00$ &$0.00$\\
%$\lambda_g$ &$-4.00$ & $-4.12$ &  &$-10.00$ &$-10.51$\\
%\end{tabular}}
%\end{table}

\section{Introduction}
\label{se:intro}

Consider $n$ realizations $(\vecx_1, \vecx_2,..., \vecx_n)$ of a $p$-dimensional random variable $\vecxc$ that follows a $G$-component finite Gaussian mixture model. The likelihood is given by
\begin{equation}
\mathcal{L}(\varthet\mid\vecx)= \prod_{i=1}^n\sum_{g=1}^G \pi_g\phi(\vecx_i\mid\vecmu_g,\matsig_g),
\end{equation} \label{eq:gauss}where $\pi_g>0$, with $\sum_{g=1}^G\pi_g=1$, are mixing proportions, $\phi(\vecx\mid\vecmu_g,\matsig_g)$ is multivariate Gaussian density with mean $\vecmu_g$ and covariance matrix $\matsig_g$, and $\varthet=(\pi_1,\ldots,\pi_G,\vecmu_1,\ldots,\vecmu_G,\matsig_1,\ldots,\matsig_G)$. A model-based clustering approach assumes that each component or some combination of components corresponds to a cluster. When fitting the model in \eqref{eq:gauss}, the main task is to decide the number of components $G$.
\cite{titterington85}, \cite{mclachan3} and \cite{mclachan} extensively reviewed mixture models, with a focus on Gaussian mixture models. \cite{fraley} presented a review of work on Gaussian mixtures with a focus on clustering, discriminant analysis, and density estimation. They discuss a family of Gaussian mixture models, which arises from the imposition of constraints upon an eigen-decomposition of the component covariance structure. The family of mixture models they discuss, known as MCLUST, is actually a subset of the Gaussian parsimonious clustering models (GPCMs) of \cite{celeux95}. When using the MCLUST models, one must choose the appropriate member of the family, i.e., the covariance structure, in addition to deciding the number of components $G$.

\cite{ghahramani97} introduced a mixture of factor analyzers model, which was further developed by \cite{tipping99b} and \cite{mclachlan00a}. Through foisting constraints on the covariance structure, \cite{mcnicholas08,mcnicholas10d} develop mixtures of factor analyzers into a family of parsimonious Gaussian mixture models (PGMMs). Now, in addition to selecting the member of the family (i.e., the covariance structure) and the number of components, one must also select the number of latent factors. Further complicating the model selection problem here is the fact that PGMMs are often applied to high-dimensional data. \cite{mcnicholas10a} explain why the PGMMs are particularly suited to the analysis of high-dimensional data: amongst the most salient points is the fact that, unlike families like MCLUST, the number of covariance parameters is linear in data dimensionality for every member of the PGMM family.

There are a number of well-known methods to select the best mixture model but the BIC remains by far-and-away the most popular. We have  
\begin{equation}
\text{BIC}=2\log \mathcal{L}(\hat\varthet \mid \vecx)-\rho\log n, \label{eq:bic}
\end{equation}
where $\hat\varthet$ is the MLE of $\varthet$, $\mathcal{L}$ is the likelihood, $\rho$ is the number of free parameters and $n$ is the number of observations. 
For a family of mixture models, the model having the maximum BIC is selected. The use of BIC is theoretically justified by a number of authors, e.g., \cite{kass1}, \cite{kass2}, and \cite{keirbin}. In particular, the BIC has some useful asymptotic properties, e.g., the criterion consistently chooses the right model under an increasing number of observations \citep{ritei}.

Nevertheless, the BIC is not without drawbacks. The criterion is derived using a Laplace approximation and its precision is influenced by the specific form of the prior density of the parameters as well as the correlation structure between observations. Recently, \cite{clyde} have rectified the problems of the marginal distribution of the parameter, caused by the Laplace approximation. In addition, \cite{fraley3}  proposed a Bayesian regularization for Gaussian mixtures. Their method assumes pre-defined priors that lead to a modified version of the BIC, using posterior modes instead of the maximum likelihood estimates (MLEs) of the parameters. The resulting method avoids degeneracies, singularities, and the problem of flat priors. However, another more serious problem has not been addressed, i.e., the problem of high-dimensional cases.

%\section{Background}\label{sec:back}
The penalty term in the BIC is $\rho\log n$, cf.\ \eqref{eq:bic}. Therefore, in a high-dimensional setting, where $p\gg n$, the penalty term dominates the likelihood and so the BIC is prone to under fitting. Parametric estimation for high-dimensional cases has been studied by a number of authors, mostly within the linear regression set-up. The celebrated LASSO method \citep{tibshirani} is perhaps the most popular among them. This method minimizes the residual sum of squares under the constraint that the sum of the absolute values of the regression coefficients is less than some constant, leading to sparse solutions of the coefficients and thus an interpretable model. In the following years, different variations of the LASSO have been proposed depending on the nature of regression and asymptotic behaviour. Some of them are the adaptive LASSO \citep{zou}, the fused LASSO \citep{tibshirani2}, and the graphical LASSO \citep{friedman}. \cite{fan} provided a theoretical discussion of variable selection via a non-concave penalized likelihood procedure where the LASSO is a special case. They also proposed that a good penalized estimation should satisfy the oracle properties, i.e., it should be consistent and the estimates should be asymptotically Gaussian. 

Following the idea of \cite{fan}, \cite{khalili} were the first to propose the use of the penalized likelihood in finite mixture of regression models, where the penalty is non-concave LASSO being a special case. They also devised a method of selecting the tuning parameter as well as conditions under which the estimation procedure would satisfy the oracle properties. Their method is especially suitable for finite mixtures of regression models, though no new model selection criterion was proposed. It should also be noted that the theoretical results regarding the asymptotic properties were somehow strange, because the authors used the same tuning parameter comparing two different estimates for a fixed cluster. \cite{chen} proposed an extended BIC for regression in high-dimensional setting. The extended BIC assumes a prior inversely proportional to the size of the assumed model instead of a flat prior. The criterion is consistent and computationally cheap. Interestingly, the authors did not propose any penalized likelihood here, instead they maximized the natural likelihood, thus using the conventional estimation procedure. The above estimation procedures, though interesting and useful, are mainly for regression-type problems, and not applicable to mixture model-based clustering and classification. Also, as the authors rightly pointed out, the approach is computationally infeasible if $p\gg n$.  Nevertheless, useful extensions can be possible. Herein, we draw upon some mathematical results from \cite{fan} and \cite{khalili}, especially on the issues of the choice of penalty and consistency.

The use of penalized likelihood in mixture model-based clustering has been proposed by \cite{pan}, where a LASSO-type penalty is applied to the likelihood. From there, they went on to propose a modified BIC which would be well-suited for high-dimensional settings. The limitation of that method is that this criterion works only for a common, diagonal component covariance matrix. Furthermore, the authors did not study the asymptotic properties, which are important in the sense that the classical LASSO method can be inconsistent \cite[cf.][]{zou}. An ideal criterion should be analytically derivable from the penalized likelihood, work well for an arbitrary model, and have some good asymptotic properties. The work presented herein attempts to address these requirements by proposing LASSO-penalized BIC (LPBIC) for model selection within high-dimensional setting for the PGMM family. 
 
While deriving the MLE of the unknown parameters, we use a penalized likelihood approach. In particular, instead of maximizing the likelihood $\mathcal{L}(\varthet \mid \vecx)$, we maximize the penalized log-likelihood 
$$\log\mathcal{L}(\varthet \mid \vecx)- \sum_{g=1}^G \pi_g \sum_{j=1}^p \varphi(\mu_{gj}).$$
We use a LASSO-like penalty for $\varphi(\mu_{gj})$. In particular, $\varphi(\mu_{gj})= n\lambda_n |\mu_{gj}|$, where $\mu_{gj}$ is the $j$th element in $\vecmu_g$ and $\lambda_n$ is the tuning parameter that depends on $n$. Though a LASSO penalty is used here, other types of non-concave penalties can also be suitable. For example, one might use the HARD penalty $\varphi(\mu_{gj})=[\lambda_n^2-\left(\sqrt{n}\mu_{gj}-\lambda_n\right)^2I\left(\sqrt{n}\mu_{gj} < \lambda_n\right)]$ or the SCAD penalty, as discussed by \cite{fan}. One problem with using such an $L_1$-norm penalty is that the oracle properties might not be satisfied fully: the estimation can be consistent but not asymptotically normal. HARD or SCAD penalties satisfy both these properties and these issues are discussed in more detail in Section \ref{asymp}. Still, however, we prefer the LASSO-type penalty because it is computationally easier due to its convexity. 
From this penalized likelihood, we derive a model selection criterion. We use a modified AECM algorithm \citep{mclachan} to estimate the parameters in the PGMM models. We show that in high-dimensional settings, our LPBIC generally outperforms the BIC for the PGMM family.

The remainder of this paper is laid out as follows. In Section~\ref{se:method}, we discuss parameter estimation under the penalized likelihood approach and derive an LPBIC. The asymptotic properties of LPBIC are discussed (Section~\ref{asymp}) and we illustrate our approach on real and simulated data (Section~\ref{se:data}). The real data considered exhibit the `small~$n$, large~$p$' property and our data analysis results are compared with the BIC. The paper concludes with a discussion (Section~\ref{sec:disc}), while the mathematical derivation of LPBIC as well as its asymptotic properties are discussed in appendices.

\section{Method}\label{se:method}
Again, suppose we observe $\vecx=\left(\vecx_1, \vecx_2,..., \vecx_n \right)$ with $f(\vecx\mid\varthet)= \sum_{g=1}^G \pi_g\phi(\vecx\mid\vecmu_g,\matsig_g)$, where $\phi(\vecx\mid\vecmu_g,\matsig_g)$ is multivariate Gaussian density with mean $\vecmu_g$ and covariance matrix $\matsig_g$.
Now, instead of maximizing the  likelihood $\mathcal{L}(\varthet \mid \vecx)$, we maximize the penalized log-likelihood 
\begin{eqnarray}
\log\mathcal{L}_{\text{pen}}(\varthet \mid \vecx)=\log\mathcal{L}(\varthet \mid \vecx)-n\lambda_n \sum_{g=1}^G \pi_g \sum_{j=1}^p|\mu_{gj}|,
\label{penlike}
\end{eqnarray}
where $\vecmu_k$ and $\lambda_n$ are defined as before.  Hereafter, we denote $\varphi(\vecmu)= \sum_{g=1}^G \pi_g \sum_{j=1}^p \varphi(\mu_{gj})$ and so
$$\log\mathcal{L}_{\text{pen}}(\varthet \mid \vecx)
=\log\mathcal{L}(\varthet \mid \vecx)-n\lambda_n\sum_{g=1}^G \pi_g \sum_{j=1}^p \varphi(\mu_{gj})
=\log\mathcal{L}(\varthet \mid \vecx)-n\lambda_n\varphi(\vecmu).$$
%By maximizing the log-likelihood with a penalty,
%there is a positive chance of having some estimated values of $\mu$ equaling 0 and thus of automatically selecting a submodel. We choose the penalty with the $k$th component to be proportional to $\pi_k$. This relates the penalty to the sample size. The virtual sample size from the kth subpopulation is proportional to $\pi_k$, and so this imposes a restriction on the number of parameters. We shall discuss the estimation of $\mathbf\mu$ under this set-up at the end.

Before going into details of parameter estimation, we make two assumptions. Firstly, as we can observe, the penalty function is non-concave and singular at the origin; it does not have second derivative at~0. We locally approximate the penalty by a quadratic function as suggested by \cite{fan}. The parameters are estimated by successive iterations. Suppose $\vecmu^{(m)}$ is the estimate of of $\vecmu$ after $m$ iterations. The penalty can be locally approximated as 
\begin{equation}
 \varphi(\vecmu)  \approx \,  
 n\lambda_n \sum_{g=1}^G \pi_g \sum_{j=1}^{p_g} \mid \mu^{(m)}_{gj} \mid+\frac{1}{2}\frac{\mbox{sign}\{\mu^{(m)}_{gj}\}}{\mu^{(m)}_{gj}}(\mu^2_{gj}-{\mu^{(m)}}^2_{gj}),\label{eq:penalty}
\end{equation}  
where $p_g$ is the number of non-zero elements in $\vecmu_g$.
We assume that the marginal distribution of the mixing proportions $\left(\pi_1,\pi_2,...,\pi_g\right)$ is uniform on the simplex and that $\vecmu_g \sim \mathcal{N}(\hat{\vecmu}_g, I(\hat{\vecmu}_g)^{-1})$, for $g=1,2,...,G$, where $\hat\vecmu_g$ is the MLE derived by maximizing  the penalized likelihood $\mathcal{L}_{\text{pen}}$ and $I(\hat\vecmu_g)$ is the unit information matrix at $\hat\vecmu_g$. 
%
%\subsection{Algorithm}
%Then we have to maximize $$L(\mu, \phi|x)+\lambda\sum_{k=1}^g\pi_k \sum_{j=1}^{p_{k}}\left[|\hat\mu_{kj}|+\frac{1}{2}\frac{\mbox{sgn}(\hat\mu_{kj})}{\hat\mu_{kj}}(\mu^2_{kj}-\hat\mu^2_{kj})\right].$$
%
%From (\ref{eq:gauss}), the complete log-likelihood function can be written as  $$\log \mathcal{L}(\varthet)=\sum_{i=1}^n\sum_{g=1}^G z_{ig}\left[\log\pi_g+\log\left\lbrace \phi\left(\vecx_i \mid \vecmu_g, \matsig_g \right)\right\rbrace \right],$$ where the 

%Note that we approximate $\varphi_{n}(\vecmu)$ by a local quadratic function as described in (\ref{eq:penalty}). 

%$$\tilde{p}_n\left(\mathbf\Theta, \mathbf{\Theta}^{(m)}\right)=n\lambda\sum_{i=1}^G \pi_k\sum_{j=1}^p\left\lbrace |\mu_{jk}^{(m)}|+\frac{\mu_{jk}^2-\mu_{jk}^{(m)^2}}{2|\mu_{jk}^{(m)} |}\right\rbrace .$$

To estimate the parameters, we use the Alternating Expectation Conditional Maximization (AECM) algorithm. There are two stages of the algorithm. At the first stage of the algorithm, when estimating $\pi_g$ and $\vecmu_g$, we define $\mathbf{z}_{i}=({z}_{i1},\ldots,{z}_{iG})$ to be indicator variables showing the component membership of the $i$th observation so that $z_{ig}=1$ if $\vecx_i$ belongs to the $g$th component and $z_{ig}=0$ otherwise. $\mathbf{z}_i$ is treated as the missing data at the first stage. Hence the expected complete data log-likelihood is $$Q( \mathbf{\pi},\vecmu)=\sum_{i=1}^n\sum_{g=1}^G \hat z_{ig}\log \pi_g + \sum_{i=1}^n\sum_{g=1}^G \hat z_{ig}\log\left\lbrace \phi\left(\vecx_i \mid \vecmu_g, \matsig_g \right)\right\rbrace -\varphi(\vecmu),$$ 
where $\hat z_{ig}={\hat\pi_g \phi(\vecx_i \mid \hat\vecmu_g,\hat\matsig_g)}/{\sum_{j=1}^G \hat\pi_j\phi(\vecx_i \mid \hat\vecmu_g,\hat\matsig_g)}.$ 
The M-step maximizes $Q$ to update the parameter estimates $\pi_g$ and $\vecmu_g$. The estimation of $\pi_g$ is complicated and has a complex analytic form. However, we have observed that in practical applications, the analytical estimate is equivalent to the estimate derived by the EM algorithm. Hence, in our analyses (Section~\ref{se:data}),  $\pi_g$ can be estimated via %For $\pi_k$,
%\begin{eqnarray}
%\frac{\partial Q}{\partial \pi_k}=\sum_{i=1}^n\left(w_{ik}^{(m)}/\pi_k-w_{ig}^{(m)}/\pi_g\right)-n\lambda\sum_{j=1}^P\left(|\mu_{kj}^{(m)}|-|\mu_{gj}^{(m)}|\right) & \mbox{for $k=1,2,...,P-1$} \nonumber.
%\end{eqnarray}
%Solving for $\mathbf\pi$ and $\mathbf\mu$ we get 
\begin{eqnarray*}
\hat\pi_g= \frac{\sum_{i=1}^n \hat z_{ig}}{n}. %\left(1+\lambda\sum_{j=1}^p\left(|\mu_{kj}^{(m)}|-|\mu_{gj}^{(m)}|\right)\right)} & \mbox{for $k=1,2,...,p-1$} \nonumber,
\end{eqnarray*}
For the mean parameters, $$\frac{\partial Q}{\partial \vecmu_{g}}=\hat\matsig_g^{-1} \sum_{i=1}^n \hat z_{ig}(\vecx_i-\hat\vecmu_g)-n\lambda_n \hat\pi_g\mbox{sign}(\hat\vecmu_g).$$
Hence
\begin{eqnarray*}
\hat{\mu}_{gj}= \begin{cases} 
\mbox{sign}(\tilde\mu_{gj})\left[ |\tilde\mu_{gj}|-\lambda_n\left(\hat\matsig_g\mathbf{1}\right)_j\right]_{+} & \text{if } \left(\hat\matsig_g\mathbf{1}\right)_j > 0,\\  \tilde\mu_{gj}& \text{otherwise}.\end{cases}
\end{eqnarray*}
where $\tilde\mu_{gj}=\sum_{i=1}^n\hat{z}_{ig}x_{ig}/\sum_{i=1}^n\hat{z}_{ig}$ is the update of $\mu_{gj}$ if no penalty term were involved, $\mathbf{1}$ is the vector with every element equal to 1, and for any $\alpha$, $\alpha_{+}=\alpha$ if $\alpha >0$ and $\alpha_{+}=0$ otherwise.  
$\hat\mu_{gj}$ is a shrunken estimate of $\mu_{gj}$ in the sense that 
$\hat\mu_{gj}=0$ if $(\mathbf{\hat\Sigma}_g\mathbf{1})_j \geq 0$ and $ \lambda_n > {\tilde\mu_{gj}}/{(\hat\matsig_g\mathbf{1})_j }.$ %\nonumber
 Otherwise, $\hat\mu_{gj}$ is obtained by shrinking the usual EM estimate $\tilde\mu_{gj}$ by the amount $ \lambda_n(\hat\matsig_g\mathbf{1})_j$ towards 0.  

At the second stage of the AECM algorithm, we take the missing data as the group labels $\mathbf{z_i}$ and the unobserved latent factors $\mathbf{u}$ to estimate the variance-covariance matrix under the PGMM set-up. The component covariance matrices $\matsig_1,\ldots,\matsig_G$ are updated as usual, depending on the family of models used; see \cite{mcnicholas08,mcnicholas10d} for details in the case of the PGMM family.
The first stage, where the $\vecmu_g$ and $\pi_g$ are estimated based on the complete data $(\vecx,\vecz)$, and the second stage, where the constituent parts of the $\matsig_g$ are estimated based on the complete data $(\vecx,\vecz,\vecu)$, are iterated until convergence. Extensive details on an AECM algorithm for fitting the members of the PGMM family are given by \cite{mclachlan00a} and \cite{mcnicholas10a}.

To derive a model selection criterion from the penalized log-likelihood, we maximize (\ref{penlike}). Using (\ref{eq:penalty}), the second term of (\ref{penlike}) becomes
%$$\lambda \sum_{k=1}^G\int \pi_k \sum_{j=1}^{P}|\mu_{kj}|\mbox{d}\pi =
$$\frac{\lambda_n}{G}\sum_{g=1}^G \sum_{j=1}^{p_{g}}\left[|\hat\mu_{gj}|+\frac{1}{2}\frac{\mbox{sign}(\hat\mu_{gj})}{\hat\mu_{gj}}(\mu^2_{gj}-\hat\mu^2_{gj})\right],$$ where $p_g$ is the number of non-zero mean components in class $g$.
Here we make an assumption that for a given model, the mixture components are chosen independently so that the parameters for any two clusters are independent. Hence, using the Weak Law of large Numbers with the BIC-type approximation to $\log\mathcal{L}(\varthet \mid \vecx)$, the penalized BIC is 
\begin{equation}
\text{LPBIC}= 2\log\mathcal{L}(\hat\varthet \mid \vecx)-\tilde \rho \log n-\frac{2n\lambda_n}{G}\sum_{g=1}^G \sum_{j=1}^{p_{g}}\left[|\hat\mu_{gj}|+\frac{\left(I(\hat{\vecmu}_g)^{-1}\right)_{jj}}{|\hat\mu_{gj}|}-\mbox{sign}\left(\hat\mu_{gj}\right)\right], \label{eq:pbic}
\end{equation} 
where $\tilde \rho$ is the number of estimated parameters which are non-zero.
Intuitively, the LPBIC further penalizes the traditional BIC by both aboslute mean and absolute coefficient of variation of the parameters. The derivation is discussed in detail in Appendix~\ref{se:app1}.

\section{Asymptotic Properties}\label{asymp}
\subsection{Properties}
The consistency of a model selection criterion is closely related to the asymptotic identifiability of the model. In general, a model $ \mathcal{G}$ with the the parameter set $\varthet$ is called identifiable if, for any two different sets of parameters $\varthet_1$ and $\varthet_2$, $$ \mathcal{G}\left(\varthet_1\right)= \mathcal{G}\left(\varthet_2\right) \quad \Longrightarrow \quad \varthet_1=\varthet_2.$$ We assume that our model satisfies the asymptotic identifiability condition. 
In the context of mixture models, a criterion is consistent if it can correctly select the number of components and the true set of parameters. If the true parameter set $\varthet_0$ is decomposed as $(\varthet_{01}, \varthet_{02})$ such that $\varthet_{02}$ contains only the zero elements, and if any estimated parameter $\hat\varthet$ that is sufficiently close to $\varthet_0$ is likewise decomposed as $(\hat\varthet_1, \hat\varthet_2)$, then in order to satisfy consistency, we should have $\text{P}(\hat\varthet_2 = \mathbf{0})\longrightarrow 1$ as $n \longrightarrow \infty$ and $\hat\varthet_1 \longrightarrow \varthet_{01} $ in probability. Thus, the criterion should choose as it would if the true number of clusters and the true parameters were known. Based on this idea, we study the consistency of LPBIC with the help of the following assumptions:
\begin{enumerate}[I]
\item Let $p = \mathcal{O}\left(n^\alpha\right)$ and $\lambda_n=o\left( \log n/n\right)$. Define an estimate $\hat\varthet$  of $\varthet$ be such that $ \mid\mid \hat\varthet-\varthet_0\mid\mid =\mathcal{O}\left(n^\kappa\right)$ for $\kappa > -\infty$.    \label{a1}
 \item Let $\varthet=\left(\theta_1,\theta_2,...,\theta_{\nu}\right)$. Then there exist finite real numbers $M_1$ and $M_2$ (possibly depending on $\kappa$) such that 
$$\sup_j\left| \frac{\partial \log\mathcal{L}(\varthet \mid \vecx)}{\partial \theta_j}\right| \leq M_1(\vecx) \quad \mbox{and} \quad \sup_{j,k}\left| \frac{\partial^2 \log\mathcal{L}(\varthet \mid \vecx)}{\partial \theta_j \partial\theta_k} \right| \leq M_2(\vecx).$$ \label{a2}
\item $I(\varthet)$ is positive-definite for all $\varthet$. \label{a4}
\end{enumerate}
 Then, under assumptions \ref{a1} to \ref{a4}, and assuming that the asymptotic identifiability condition is satisfied, we state the following theorem. The proof is given in Appendix~\ref{se:app2}.
\begin{theorem}
 If $\kappa <\min\left[0, {(\alpha-1)}/2\right]$, then the LPBIC chooses the number of components and set of parameters as it would choose if $\varthet_0$ were known as $n \longrightarrow \infty$. In other words, under the condition $\kappa <\min\left(0, {\alpha-1}/2\right)$, if there exists an estimate $\tilde\varthet$ such that $||\tilde\varthet-\varthet_0|| = \mathcal{O}\left(n^\kappa\right)$ and  $\text{LPBIC}(\tilde\varthet)\geq\text{LPBIC}(\varthet)$ for all $\varthet$  such that $||\varthet-\varthet_0|| = \mathcal{O}\left(n^\kappa\right)$, then
 \begin{pta}
 P$(\tilde\varthet_2= \mathbf{0})\longrightarrow 1$ as $n \longrightarrow \infty $, and 
 \end{pta}
 \begin{ptb}
 $\tilde\varthet_1 \longrightarrow \varthet_{01} $ in probability as $n \longrightarrow \infty$.
 \end{ptb}
\end{theorem}

We prove only Part a with some of the arguments proposed by \cite{khalili} for an FMR setting. The method is modified for mixture models with high-dimensional set-up. Part b of the theorem can be proved exactly by the method described in \cite{fan}. 
To prove Part b, we need $\sqrt{n}\lambda_n \rightarrow 0$, as $n \rightarrow \infty$ which is satisfied by Assumption \ref{a1}. This is particularly important because LASSO-type penalties do not satisfy the oracle property, i.e., they do not ensure that a $\sqrt{n}$-consistent MLE of $\theta$ exists which satisfies Part a and Part b. This is because the existence of a $\sqrt{n}$-consistent MLE requires that $\sqrt{n}\lambda_n \longrightarrow \infty$ and the consistency of $\hat\vartheta_1$ needs that  $\sqrt{n}\lambda_n \longrightarrow 0$. Hence, under a tighter assumption, we show that if such an estimator exists, then it satisfies consistency. Other non-concave penalties like SCAD or HARD, however, can satisfy the oracle property with a proper choice of the tuning parameter.

\subsection{Choice of $\lambda_n$}
Generally the tuning parameters are chosen by cross-validation \citep{stone} or generalized cross-validation \citep{craven}. We should remember that $\lambda_n$ depends on $n$. To satisfy the asymptotic properties, we require $\lambda=o\left(\log n/n\right)$.  \cite{khalili} derived a component-wise deviance-based GCV with the above conditions in order to estimate $\lambda$. The method, though originally used in regression, also serves well for mixture models. The present paper takes the working sequence $\lambda_n= 1/p$ and studies the behaviour of the LPBIC. The methods proposed by \cite{khalili}, modified for a mixture model, are also considered and provide a range for the values of $\lambda_n$. It is observed that for moderately large $n$ ($n \geq 50$), $\lambda_n= 1/p$ falls into that range. For our data analysis (Section~\ref{se:data}), we studied the behaviour of LPBIC for different values of $\lambda_n$ within that range. For illustration, though, a single $\lambda_n$ is chosen because the behaviour of the LPBIC is uniform over different $\lambda_n$ values within that range.

\section{Data Analysis}\label{se:data}
\subsection{Overview}
We analyze two data sets and compare the results using the PBIC to those with the BIC for the PGMM family. The first one is a high-dimensional simulated data set and the second one is a real high-dimensional data set. Although run as cluster analyses, the true group memberships are known in each case and we use the adjusted Rand index \citep[ARI:][]{rand,hubert} to reflect classification agreement. A value of $1$ indicates perfect agreement and a value of $0$ would be expected under random classification.

\subsection{Simulated Data}
We generate a simulated $p$-dimensional Gaussian data set consisting of three groups.
 We set $\vecmu_1=-5.5\mathbf{1}$, $\matsig_1$ isotropic; $\vecmu_2=2\mathbf{1}$, $\matsig_2$ diagonal; and $\vecmu_3=3\mathbf{1}$, $\matsig_3$ full, with  $n_1=40$, $n_2=30$, $n_3=30$.  We ran simulations for $p\in\{100, 250,500\}$. LPBIC values are observed for each member of the PGMM family for $G=1,\ldots,4$ and $q=1,2,3$. The results (Table~\ref{tab:sim1p}) show that the PBIC consistently chooses $G=3$ as $p$ gets larger but that the BIC fails in higher dimensions, choosing a $G=2$ component model. The associated ARI values (Table~\ref{tab:sim1p}) confirm that the models selected by the PBIC capture the underlying group structure better than those chosen by the BIC, especially in higher dimensions.  %the estimates of $\lambda_n$ are .01, .006 and .002 for $p=100$, 250 and 500, respectively. 
\begin{table}[!h]
\caption{\label{tab:sim1p} Best model chosen by PBIC and BIC for high-dimensional simulated data.}
\centering
\begin{tabular*}{0.95\textwidth}{@{\extracolsep{\fill}}llllllllll}
\hline
   & \multicolumn{4}{c}{LPBIC} && \multicolumn{4}{c}{BIC} \\
\cline{2-5} \cline{7-10}
 & $G$ & $q$ & Model & ARI && $G$ & $q$ & Model & ARI\\
\cline{2-5} \cline{7-10}
$p=100$ & $3$ & $3$ & CUC&$0.88$ && $3$& $3$ & CUC &$0.86$\\ 
$p=250$  & $3$ & $2$ & CUC & $0.82$ && $2$ & $1$ & CCC & $0.62$ \\
$p=500$  & $3$ & $3$ & CUC & $0.97$ && $2$ & $1$ & CCC & $0.49$\\
\hline\end{tabular*}
\end{table}

The effect of increasing dimension on the performance of the BIC is clear: the BIC chooses fewer mixture components and latent factors, as well as a more parsimonious covariance structure. The LPBIC, however, chooses the same number of components and the same covariance structure each time, and the number of factors does not decrease with $p$. 

\begin{figure}[!h]
\begin{center}
\begin{tabular}{cc}
\includegraphics[width=0.45\textwidth]{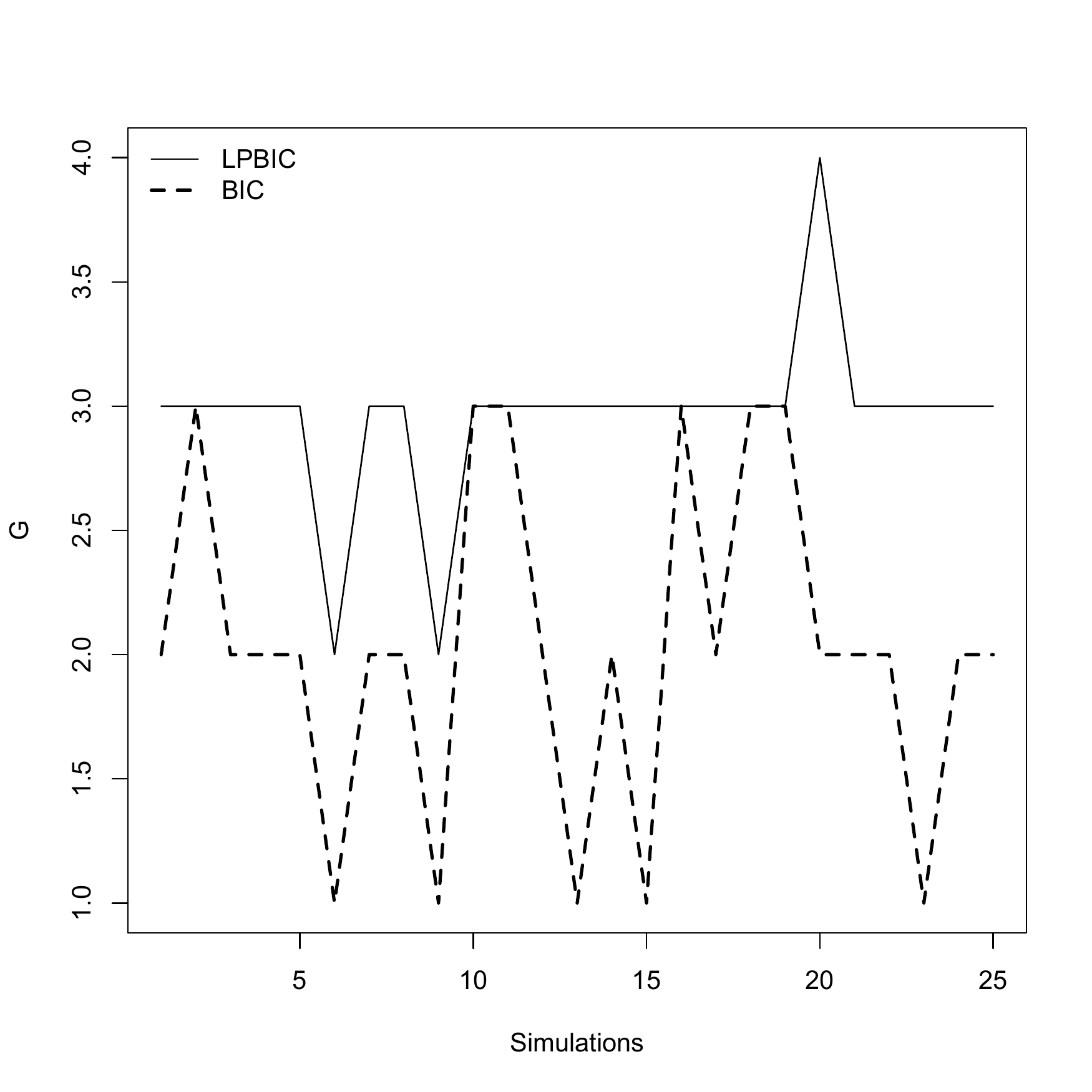} & \qquad \includegraphics[width=0.45\textwidth]{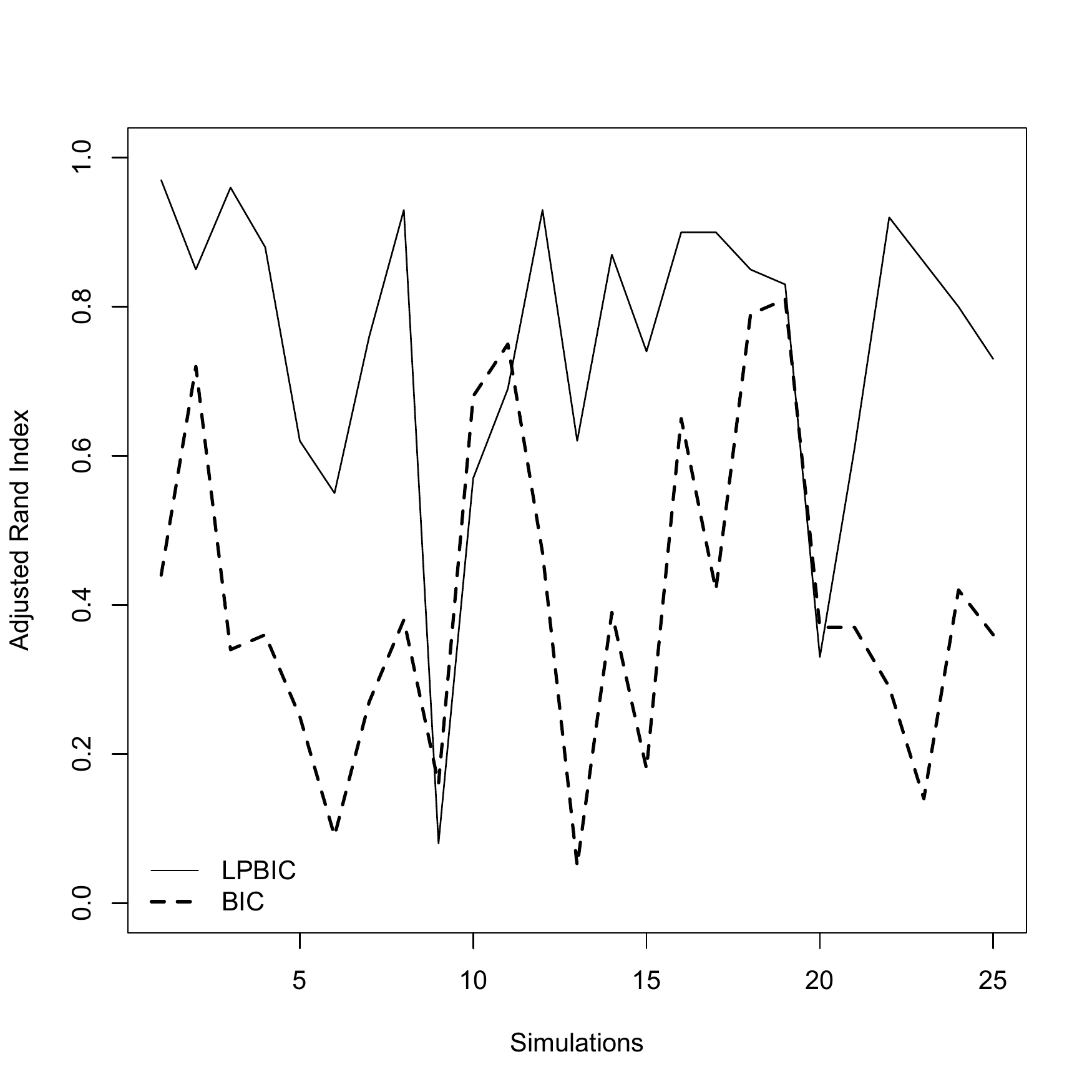}
\end{tabular}
\end{center}
\vspace{-0.2in}
\caption{Plot of the performance of LPBIC and BIC for 25 simulations. The left-hand plot shows the selection of number of components by the BIC and LPBIC. The right-hand plot shows the ARIs of the models selected by LPBIC and BIC.}\label{fi:sim1}
\end{figure}

Next, we generate 25 simulations of the $p=500$ dimensional data and study the behaviour of BIC and LPBIC for selecting~$G$ and for clustering performance (i.e., ARI). The results (Figure \ref{fi:sim1}) show that LPBIC correctly chooses the number of components ($G=3$) 23 times but the BIC only selects $G=3$ four times out of 25. As expected, the BIC tends to choose too few components. The ARIs for models selected using the LPBIC are higher than those selected using the BIC. Out of 25 simulations, the ARI with the LPBIC is higher than that for the BIC in 21 cases, illustrating generally superior clustering performance.

% Interestingly, BIC chooses $q=1$ for as many as 21 cases, all of which are with the model CCC. This indicates that with high-dimension, BIC has a tendency to undertone the covariance structure and latent factor. Due to complex inter-relations between $G$, $q$ and the covariance structure, it is difficult to decide which of the above three failures is the gravest. But overall we can say that BIC portrays a tendency to underestimate the model structure for high-dimensional cases. LPBIC chooses the model CUC 22 times, and the model CCU 2 times. For the 9th simulation, LPBIC chooses the model CCC. As apparent from Figure \ref{fi:sim1}, the 9th simulation may be an anomaly because both the criteria failed to choose the correct number of components, and the adjust Rand indices of both are unusually low.

\subsection{Leukaemia data}

\cite{golub} presented data on two forms of acute leukaemia: acute lymphoblastic leukaemia (ALL) and acute myeloid leukaemia (AML). Affymetrix arrays were used to collect measurements for 7,129 genes on 72 tissues. There were a total of 47 ALL tissues and 25 with AML. \cite{mclachlan02} reduced the data set as follows:
\begin{enumerate}
\item Genes with expression falling outside the interval $(100, 16000)$ are removed.
\item Genes with expression satisfying max/min $\leq 5$ or max-min $\leq 500$ are removed.
\end{enumerate}
\cite{mcnicholas10d} further reduced the number of genes to 2,030 by applying the {\tt select-genes} software \citep[cf.][]{mclachlan02}. We analyze these 2,030 genes using 20 different random starts for the initial $\hat z_{ig}$. We run our approach for $G\in\{1,2\}$ and $q=1,\ldots,6$.
\begin{table}[!h]
\caption{Comparison of the performance of LPBIC and BIC for PGMM model selection for the leukaemia data.}
\centering
\begin{tabular*}{0.95\textwidth}{@{\extracolsep{\fill}}lllllr}
\hline
  & Value & $G$ & $q$ & Model & ARI \\
\hline 
BIC & $-400394$ & $1$ & $2$ & CCU &  0.29\\ 
LPBIC & $-391023$ & $2$ & $1$ & CUC & 0.47 \\
%LPBIC (second best)&$-391152$ & $2$ & $2$ & CCU & 0.51 \\
\hline 
\end{tabular*} \label{tab:luke1}
\end{table}

Summaries of the models selected by the LPBIC and the BIC, respectively, are given in Table~\ref{tab:luke1}. The BIC chooses a CCU model with $G=1$ component and $q=2$ factors. The LPBIC chooses a CUC model with $G=2$ components and $q=1$ factors. The ARI of the model chosen using LPBIC ($0.47$) is greater than that for the model chosen using the BIC ($0.29$). %Interestingly, the ARI associated with the second best model (CCU with $G=2$ and $q=2$), in terms of LPBIC, is better than that for the best model (cf.\ Table~\ref{tab:luke1}). This improvement in classification amounts to two fewer misclassifications: 9 as opposed to 11 for the best model selected with LPBIC (Table~\ref{tab:luke2}).
The model selected using the LPBIC misclassifies eleven of the $72$ samples (Table~\ref{tab:luke2}).
\begin{table}[!h]
\caption{\label{tab:luke2} Classification table of the best model chosen by LPBIC.}
\centering
\begin{tabular*}{0.95\textwidth}{@{\extracolsep{\fill}}lrr}
\hline
   & 1 & 2  \\
\hline 
ALL & 39 & 3 \\ 
AML  & 8 & 22 \\
\hline
\end{tabular*}
\end{table}

\section{Discussion}\label{sec:disc}

The paper proposes a LPBIC through a penalized likelihood-based approach in the context of parsimonious Gaussian mixture model selection. The approach is mainly intended for the high-dimensional setting, where the BIC has some unattractive problems due to an `exploding' penalty term for high-dimensional data. Our LPBIC approach does not use the total number of independent parameters to be estimated in its penalty term but, rather, the total number of independent non-zero parameters to be estimated. This has some advantages. Because the likelihood is penalized by a tuning parameter, many of the mean components become $0$, thereby reducing the number of independent estimable parameters. The loss of information due to penalizing the likelihood is somehow compensated for by both absolute mean and absolute coefficient of variation of the mean parameters. 

The choice of tuning parameters is an important aspect in this scenario because no theoretical result exists which specifies the best choice. Recently, \cite{wang1,wang2} proposed some interesting mathematical methods of choosing the tuning parameters without requiring cross-validation. However, their method is most suitable in low-dimensional settings. Herein, we followed an approach close to the one proposed by \cite{fan}, though careful modifications have been taken to preserve the asymptotic properties, accounting for the nature of the data. 

Our method seems consistent in choosing the right number of clusters for high-dimensional data, as shown through the analysis of real and simulated data. Our analyses suggest that the LPBIC is an improvement over the BIC in the high-dimensional setting. What we lose is the oracle property, because the LASSO may fail to satisfy the consistency, sparsity and asymptotic normality all at the same time. But the LASSO has some computational advantages because of convexity and hence it is preferred over other non-concave penalties.

Of course, the LPBIC is not without its issues. One problem arises by locally approximating the penalty function: if an estimator is shrunken, it stays at 0. Another arises if the initial domain of the estimates does not contain the posterior mode, or even if the posterior mode lies at the boundary of the domain. This second problem, which will lead to failure, is a general problem with the EM algorithm.

Future work will focus on the use of penalties that lead to consistent model selection criteria. We are in particular interested in the adaptive LASSO which leads to the oracle properties. We shall also study the penalization of the variance parameters as it will generate greater parsimony. 

\section*{Acknowledgements}
This work was presented at the Statistical Society of Canada annual meeting in Guelph, Canada (June~2012) and at the MBC$^2$ meeting in Catania, Italy (September~2012). The authors wish to thank the numerous people who provided helpful comments and feedback at these meetings. This work was supported by a Collaborative Research and Development grant from the Natural Sciences and Engineering Research Council of Canada, a grant-in-aid from Compusense Inc., a Collaborative Research grant from the Ontario Centres of Excellence, and an Early Researcher Award from the Ontario Ministry of Research and Innovation.   

\bibliographystyle{chicago}
\bibliography{biblio}

\appendix
\section{Derivation of LPBIC}
\label{se:app1}
To derive the LPBIC, we closely follow the derivation of the usual BIC. We have to maximize (\ref{penlike}). Using (\ref{eq:penalty}), the second term becomes 
%$$\lambda \sum_{k=1}^G\int \pi_k \sum_{j=1}^{P}|\mu_{kj}|\mbox{d}\pi =
$$n\lambda_n\sum_{g=1}^G\int \pi_g \sum_{j=1}^{p}|\mu_{gj}|\mbox{d}\mathbf\pi_g=
n\frac{\lambda_n}{G}\sum_{g=1}^G \sum_{j=1}^{p_{g}}\left[|\hat\mu_{gj}|+\frac{1}{2}\frac{\mbox{sign}(\hat\mu_{gj})}{\hat\mu_{gj}}(\mu^2_{gj}-\hat\mu^2_{gj})\right],$$ where $p_g$ is the number of non-zero mean components in class~$g$.
Under the assumption made in Section~ \ref{se:method}, $\vecmu_g$ is at most $p_g$ dependent, and the Weak Law of Large Numbers holds. In a large-$p$ setting, $\sum_{g=1}^G p_g$ is a large number and so 
$\sum_{g=1}^G \sum_{j=1}^{p_{g}}\left(\mu^2_{gj}-\hat\mu^2_{gj}\right)/{\sum_{g=1}^G p_g}  \overset{P}{\longrightarrow}\sum_{g=1}^G \sum_{j=1}^{p_{g}}\left(I(\hat{\mathbf\mu}_g)^{-1}\right)_{jj}/{\sum_{g=1}^G p_g}.$
Thus the second term becomes $$\frac{n\lambda}{G}\sum_{k=1}^G \sum_{j=1}^{p_g}\left(|\hat\mu_{gj}|+\frac{\left(I(\hat{\mathbf\mu}_g)^{-1}\right)_{jj}}{|\hat\mu_{gj}|}\right).$$

The first term, using Taylor's expansion, is 
\begin{equation*}\begin{split}
\int \text{exp}&\left[\log \mathcal{L}(\varthet \mid \vecx)\mathcal{G}(\varthet)\right]\mbox{d}\mathbf\Theta\\&= \int \text{exp}\left[\log \mathcal{L}(\hat{\vecthet} \mid \vecx)\mathcal{G}(\hat{\varthet})+ (\varthet-\hat{\varthet})\frac{\partial \log \mathcal{L}(\varthet)\mathcal{G}(\varthet)}{\partial{\varthet}}-\frac{1}{2}(\varthet-\hat{\varthet})^T \mathcal{H}_{\hat{\varthet}}(\varthet-\hat{\varthet})\right]\mbox{d}\varthet,
\end{split}\end{equation*}
where $\mathcal{H}$ is the second derivative matrix of $\log \mathcal{L}(\varthet)\mathcal{G}(\varthet)$. Because $\hat{\varthet}$ is derived maximizing the penalized likelihood, the second term within the integral becomes $(\varthet-\hat{\varthet}) \partial\varphi_n(\vecmu)/{\partial{\varthet}}$, where $\varphi_n(\vecmu)$ is the LASSO penalty function. Using (\ref{eq:penalty}), the mean-value theorem and the fact that the $\varthet$ values are close to $\hat{\varthet}$, the second term within the integral is ${n\lambda_n}/G\sum_{g=1}^G \sum_{j=1}^{p_{g}}\mbox{sign}(\mu_{gj})$.

The third term within the integral similarly becomes $1/2 (\tilde{\varthet}-\hat{\tilde{\varthet}})'\mathcal{H}_{\hat{\tilde{\varthet}}}(\tilde{\varthet}-\hat{\tilde{\varthet}})$, where $\tilde{\varthet}$ is the set of non-zero parameters and $\hat{\tilde{\varthet}}$ is their estimate. Using Laplace approximation on $\mathcal{H}$ and applying the Weak Law of Large Numbers, as in the usual BIC, we arrive at $\log \mathcal{L}(\hat{\varthet} \mid \vecx)-{1}/{2}\tilde \rho \log n$, where $\tilde \rho= \mbox{dim}(\hat{\tilde{\varthet}}) $. This, combined with the second term of (\ref{penlike}), gives (\ref{eq:pbic}).

\section{Proof of the Asymptotic Property of LPBIC}
\label{se:app2}

First, suppose the true number of clusters $G$ is known with the corresponding parameter $\varthet$. Let the true parameter be $\varthet_0$. Let $\hat\varthet$ be an arbitrary estimate of $\varthet$.  Let $\tilde\rho_0$ and $\tilde\rho_1$ be the corresponding number of non-zero parameters and $\lambda_n^{(0)}$ and $\lambda_n^{(1)}$ be the corresponding tuning parameters. 
We first prove that, for an arbitrary estimate $\hat\varthet$ satisfying $\mid\mid \hat\varthet -\varthet_0 \mid\mid = \mathcal{O}\left( n^\kappa \right)$, $\text{LPBIC}(\hat\varthet_{1},\hat\varthet_{2})- \text{LPBIC}(\hat\varthet_{1}, \mathbf{0})\leq 0$ as $n \longrightarrow \infty$. We note that 
 $$\text{LPBIC}(\hat\varthet_{1},\hat\varthet_{2})- \text{LPBIC}(\hat\varthet_{1}, \mathbf{0}) = 2\l(\hat\varthet_{1}, \hat\varthet_{2} \mid \vecx)-2l(\hat\varthet_{1},\mathbf{0} \mid \vecx) -\left[ \Lambda(\hat\varthet_{1}, \hat\varthet_{2})-\Lambda(\hat\varthet_{1},\mathbf{0})\right],$$ 
 where $l=\log\mathcal{L}$ and $\Lambda$ is the penalty part of LPBIC. Using the mean-value theorem, $$
l(\hat\varthet_{1}, \hat\varthet_{2} \mid \vecx)-l(\hat\varthet_{1}, \mathbf{0}\mid \vecx)
=\left[\frac{\partial l(\hat\varthet_{1}, \mathbf{\xi})}{\partial \varthet_{2}}\right]'\hat\varthet_{2},$$
where $\mid\mid \mathbf{\xi} \mid\mid \leq \mid\mid \hat\varthet_{2} \mid\mid = \mathcal{O}\left(n^{\kappa}\right).$ Also,  
\begin{equation}\begin{split}\label{eqn:b1}
\left| \left| \frac{\partial l(\hat\varthet_{1}, \mathbf{\xi})}{\partial \varthet_{2}}-
\frac{\partial l\left(\varthet_{0}, \mathbf{0}\right)}{\partial \varthet_{2}} \right|\right| \leq & \left|\left| \frac{\partial l(\hat\varthet_{1}, \mathbf{\xi})}{\partial \varthet_{2}}-\frac{\partial l(\hat\varthet_{1}, \mathbf{0})}{\partial \varthet_{2}} \right|\right|+\left|\left| \frac{\partial l(\hat\varthet_{1}, \mathbf{0})}{\partial \varthet_{2}}-\frac{\partial l\left(\varthet_{0}, \mathbf{0}\right)}{\partial \varthet_{2}} \right|\right|\\
\leq & \sum_{i=1}^n M_2(z_i) \left[\left|\left|\mathbf{\xi}\right|\right|+ \left|\left| \hat\varthet_{1}-\varthet_0\right|\right|\right] = \left[\left|\left|\mathbf{\xi}\right|\right|+ \left|\left| \hat\varthet_{1}-\varthet_0\right|\right|\right] \mathcal{O}\left(n\right)=\mathcal{O}\left(n^{\kappa+1}\right)
\end{split}\end{equation} 
from Assumption~\ref{a2}.
Also, from the last part of the first line of \eqref{eqn:b1}, which is of order $\mathcal{O}\left(n^{\kappa+1}\right)$, we can conclude that $\partial l\left(\varthet_{0}, \mathbf{0}\right)/{\partial \varthet_{2}}$ is of order $\mathcal{O}\left(n^{\kappa+1}\right)$, as is $\partial l(\hat\varthet_{1}, \mathbf{\xi})/{\partial \varthet_{2}}$. Therefore, from these order assessments, we conclude that $$l\left(\tilde\varthet_{1}, \tilde\varthet_{2}\right)-l\left(\tilde\varthet_{1}, \mathbf{0}\right) = \mathcal{O}\left(n^{\kappa+1}\right)\sum_{g=1}^G\sum_{j=p_g+1}^p \hat\mu_{gj},$$ where $p_g$ is defined as in (\ref{eq:penalty}).

For the part $\Lambda(\hat\varthet_{1}, \varthet_{2})-\Lambda(\hat\varthet_{1}, \mathbf{0})$, note that 
$$\frac{2n\lambda_n}{G}\sum_{g=1}^G \sum_{j=1}^{p_{k}}\left[|\hat\mu_{gj}|+\frac{\left(I(\hat{\vecmu}_g)^{-1}\right)_{jj}}{|\hat\mu_{gj}|}-\mbox{sign}\left(\tilde\mu_{gj}\right)\right] = \mathcal{O}\left(n^{\alpha+1}\right)\lambda_n$$ because the summation part is some constant times $p=\mathcal{O}\left(n^\alpha\right)$, using Assumption~\ref{a1}. We also have $(\tilde\rho_1-\tilde\rho_0)\log n= \sum_{g=1}^G\left(p-p_g\right)\log n=\mathcal{O}\left(n^{\alpha}\right)\log n$. Hence, 
$$\text{LPBIC}(\hat\varthet_{1}, \hat\varthet_{2})- \text{LPBIC}(\hat\varthet_{1}, \mathbf{0})=\mathcal{O}(n^{\kappa+1})\sum_{g=1}^G\sum_{j=p_g+1}^p \hat\mu_{gj}-\mathcal{O}(n^{\alpha})\log n-(\lambda_n^{(1)}-\lambda_n^{(0)})\mathcal{O}(n^{\alpha+1}).$$ 
The first term of the above expression is $\mathcal{O}\left(n^{\kappa+1}\right)\sum_{g=1}^G\sum_{j=p_g+1}^p \hat\mu_{gj} =   \mathcal{O}\left(n^{2\kappa+1}\right)$.
Using Assumption \ref{a1}, i.e., that $\lambda_n =\emph {o}\left(\log n/n\right)$, and by order comparison, we can conclude that the leading terms in the above expression are $\mathcal{O}\left(n^{2\kappa+1}\right)$ and $\mathcal{O}\left(n^\alpha\right)\log n$. Because $\alpha > 2\kappa+1$, $\text{LPBIC}\left(\hat\varthet_{1}, \hat\varthet_{2} \right)- \text{LPBIC}\left(\hat\varthet_{1}, \mathbf{0}\right) \leq 0$ as $n \longrightarrow \infty$. 

Now, let $\tilde\varthet= \left(\tilde\varthet_1, \tilde\varthet_2 \right)$ be an estimate of $\varthet$ such that $(\tilde\varthet_1, \mathbf{0})$ is a maximizer of LPBIC$(\varthet_1, \mathbf{0})$ satisfying $\mid\mid \tilde\varthet-\varthet_0 \mid\mid = \mathcal{O}\left(n^\kappa\right)$. It suffices to show that in the neighbourhood $\mid\mid \varthet-\varthet_0 \mid\mid = \mathcal{O}\left(n^\kappa\right)$, $\text{LPBIC}\left(\varthet_1, \varthet_2\right)-\text{LPBIC}\left(\tilde\varthet_1, \mathbf{0}\right) < 0$ with probability tending to 1 as $n \rightarrow \infty$. We note that
$$\text{LPBIC}(\varthet_1, \varthet_2) - \text{LPBIC}(\tilde{\varthet}_1, \mathbf{0}) =
 [\text{LPBIC}(\varthet_1, \varthet_2)-\text{LPBIC}(\varthet_1, \mathbf{0})] + [\text{LPBIC}(\varthet_1, \mathbf{0})-  \text{LPBIC}(\tilde{\varthet}_1, \mathbf{0})],$$
where $\text{LPBIC}(\varthet_1, \varthet_2)-\text{LPBIC}(\varthet_1, \mathbf{0})\leq 0$ with probability tending to 1 (by the previous result) and $\text{LPBIC}(\varthet_1, \mathbf{0})- \text{LPBIC}(\tilde{\varthet}_1, \mathbf{0})\leq 0$ with probability tending to 1 since $(\tilde\varthet_1, \mathbf{0})$ is a maximizer of LPBIC$(\varthet_1, \mathbf{0})$. Thus $(\tilde\varthet_1, \mathbf{0})$ maximizes $\text{LPBIC}(\varthet_1, \varthet_2)$ with probability tending to 1 as $n \rightarrow \infty$. Hence we conclude that $\text{P}\left(\tilde\varthet_2 = \mathbf{0}\right) \longrightarrow 1$ as $n \rightarrow \infty$. Hence the proof.

The case of unknown clusters can be similarly proved. If the estimated number of components is $G_1$ and the true number is $G$,  then the estimated parameter corresponding to $G_1$ is, say, $\hat\varthet$. We can again decompose $\hat\varthet$ as $(\hat\varthet_1, \hat\varthet_2)$ and similarly show that $\hat\varthet_2 \longrightarrow 0$ in probability. Here $\hat\varthet_1$ comprises of the clusters belonging to $\varthet_0$.

\end{document}